\documentclass[runningheads]{llncs}
\usepackage[T1]{fontenc}

\usepackage[english]{babel}
\usepackage[T1]{fontenc}
\usepackage[utf8]{inputenc}
\usepackage{amsfonts,amsmath}
\usepackage{algorithm}%
\usepackage{algorithmicx}%
\usepackage{algpseudocode}%
\usepackage{booktabs}
\usepackage{mathdots}
\usepackage{graphicx}
\usepackage{subcaption}
\usepackage{hyperref}
\usepackage{paralist}
\usepackage{enumitem}
\usepackage{array}
\usepackage{tabu}
\usepackage{wrapfig}
\usepackage{float} 
\usepackage{listings}
\usepackage{pgfplots}
\usepackage{tikz}
\usetikzlibrary{automata,arrows,positioning,calc}
\usepgfplotslibrary{fillbetween}
\usepackage{longtable}
\usepackage{adjustbox}
\usepackage{forest}
\usepackage{ragged2e}
\usepackage{makecell}

\usepackage{todonotes}

\usepackage{caption}
\usepackage{multirow,multicol}

\usepackage{xspace}

\definecolor{codegreen}{rgb}{0,0.6,0}
\definecolor{codegray}{rgb}{0.5,0.5,0.5}
\definecolor{codepurple}{rgb}{0.58,0,0.82}
\definecolor{backcolour}{rgb}{0.95,0.95,0.92}
\lstdefinestyle{mystyle}{
    commentstyle=\color{codegreen},
    keywordstyle=\color{magenta},
    numberstyle=\tiny\color{codegray},
    stringstyle=\color{codepurple},
    basicstyle=\ttfamily\footnotesize,
    breakatwhitespace=false,         
    breaklines=true,                 
    captionpos=b,                    
    keepspaces=true,                 
    showspaces=false,                
    showstringspaces=false,
    showtabs=false,                  
    tabsize=2,
    aboveskip=20pt,
    belowskip=0pt
}
\usetikzlibrary{chains}

\newcommand{\snode}[2]{\node(#1)[item]{\ensuremath{#2}}}

\newcommand{\probtest}{\textit{ProbTest}\xspace}

\newcommand{\sample}{\smash{\ensuremath{\overset{\$}{\leftarrow}}}}

\newfloat{lstfloat}{htbp}{lop}
\floatname{lstfloat}{Listing}

\usepackage{caption}
\usepackage{cleveref}
\usepackage{makecell}
\newcolumntype{P}[1]{>{\centering\arraybackslash}p{#1}}

\lstdefinestyle{mystyle}{
	backgroundcolor = \color{white},
	commentstyle = \color{codegreen},
	keywordstyle = \color{magenta},
	numberstyle = \tiny\color{codegray},
	stringstyle = \color{codepurple},
	basicstyle = \ttfamily\small,
	breakatwhitespace = false,
	breaklines = true,
	captionpos = b,
	keepspaces = true,
        columns=flexible,
	numbersep = 3pt,
	showspaces = false,
	showstringspaces = false,
	showtabs = false,
	tabsize = 2
}

\newenvironment{proof*}[1][Proof]{%
  \par\noindent\emph{#1.}\ }{\hfill$\square$\par}

\begin{document}

\title{ProbTest: Unit Testing for Probabilistic Programs\\(Extended Version)}

\author{
  Katrine Christensen \and
  Mahsa Varshosaz \and
  Ra{\'u}l Pardo}
\authorrunning{K. Christensen et al.}
\institute{
  IT University of Copenhagen, Denmark
}

\maketitle

\begin{abstract}

Testing probabilistic programs is non-trivial due to their stochastic nature.
Given an input, the program may produce different outcomes depending on the underlying stochastic choices in the program. 
This means testing the expected outcomes of probabilistic programs requires repeated test executions unlike deterministic programs where a single execution may suffice for each test input.
This raises the following question: how many times should we run a probabilistic program to effectively test it?
This work proposes a novel black-box unit testing method, \probtest, for testing the outcomes of probabilistic programs.
Our method is founded on the theory surrounding a well-known combinatorial problem, the \emph{coupon collector's problem}.
Using this method, developers can write unit tests as usual without extra effort while the number of required test executions is determined automatically with statistical guarantees for the results. 
We implement \probtest as a plug-in for PyTest, a well-known unit testing tool for python programs. Using this plug-in, developers can write unit tests similar to any other Python program and the necessary test executions are handled automatically.
We evaluate the method on case studies from the Gymnasium reinforcement learning library and a randomized data structure. 

\end{abstract}

\section{Introduction}
Probabilistic programs are programs with stochastic behavior.
They have been long used in several domains such as the implementation of randomized algorithms~\cite{RandomizedAlgorithms,FoundationsOfPP,PierroW97,morgan1999pgcl}, and are at the core of widespread machine learning systems~\cite{10.5555/1162264,pml2Book,sutton1998reinforcement,statrethinkingbook}.
Probabilistic programs may return different outcomes for the same input due to their stochastic behavior.
Hence, the output of a probabilistic program induces a probability distribution over possible program outcomes. 
This stochastic behavior makes testing probabilistic programs notoriously difficult.
\looseness -1

As an example, consider the program in Listing~\ref{lst:bounded_random_walk}.
It implements a one dimensional discrete random walk.
Although this program looks simple, random walks are the core of multiple probabilistic inference methods such as Markov Chain Monte Carlo (MCMC)~\cite{brooks2011handbook}.
Consider a developer whose task is to test whether it is possible for this program to assign a value larger than a boundary $b$ to the (returned) variable \lstinline|pos|.
This property may be useful, for instance, to detect integer overflow bugs.
Evidently, to check this property the developer must run the program repeatedly.
However, this leads to the following question: \emph{``How many times is it necessary to run the program to detect a violation of this property?''}
Answering this question is not a trivial task.
The challenge illustrated in this simple example transcends to many domains and large software systems with probabilistic behavior.
For instance, how many times should we execute a test to assert a property in a randomized data structure, in a probabilistic protocol, or in a policy generated by a Reinforcement Learning (RL) algorithm?
\looseness -1

\begin{lstlisting}[
  float=tp, language=python, 
  caption={Python implementation of a discrete one dimensional random walk of \lstinline|n| steps.\looseness -1}, 
  label={lst:bounded_random_walk}]
def n_random_walk(n: int):
    pos = 0 # init_pos
    X = bernoulli(1/2) # from scipy.stats
    for _ in range(n):
        x_sample = X.rvs()
        if x_sample == 1 
            pos = pos + 1
        if x_sample == 0
            pos = pos - 1
    return pos
\end{lstlisting}

Existing work in this field tackles this problem by using a model of the program~\cite{pp_coverage_modelbased,StatisticalModelChecking,IntroductionToStatisticsAndDataAnalysis,DBLP:conf/cav/KwiatkowskaNP11,DBLP:journals/sttt/HenselJKQV22}, such as a Markov Decision Process~\cite{DBLP:books/daglib/0020348}. 
However, this requires white-box access to the program and its libraries, which is not always available.
It also requires to generate a model from the program, which creates a gap between the actual program and the model.
Other works \cite{testing_pp_statistical_hypothesis,DBLP:conf/sigsoft/DuttaLHM18} propose to find bugs by performing statistical tests between the output distribution of the program and a distribution of the target property.
Unfortunately, applying and interpreting the results of these methods often requires expertise in statistics.
For example, testers must be able to provide a distribution for the target property; or at least its mean and variance.
In practice, testers resort to running probabilistic programs an arbitrary number of times; hoping to trigger existing bugs, but without any guarantees.

In this work, we present a novel black-box unit testing method with statistical guarantees, named \probtest, to effectively test probabilistic programs. 
As it is black-box, the tester does not need to know the internal structure of the program. 
Testers specify an assertion (as in regular unit testing), and the probability of violating the assertion, i.e., the expected failure probability. 
An example for the property and program above is \emph{``The probability that the program in listing~\ref{lst:bounded_random_walk} returns a value greater than a boundary $b$ is at most 10\%''}.
\probtest\ is proven to ensure that, if a bug exists in the program, an assertion failure is triggered with probability \(1-\epsilon\); where \(\epsilon>0\) is an error probability selected by the tester.
To this end, \probtest\ automatically (and efficiently) computes the number of unit test executions ($k$) needed to trigger an assertion violation with probability \(1-\epsilon\).
Thus, our method guarantees that if all $k$ runs of the test pass, the property holds with probability $1-\epsilon$. 
If a test run fails, the property does not hold for the program, and we have discovered a bug.
As is a unit testing method, the complexity of properties that \probtest handles is naturally lower than traditional verification techniques such as statistical model-checking.
However, the number of executions \(k\) required by \probtest\ can be orders of magnitude lower than for statistical model-checking or testing methods based on statistical tests (cf.~\cref{chapter:related_work}).
The underlying theory for \probtest\ is based on the well-known \emph{coupon collector's problem}~\cite{ccp_generalisedCouponCollectorProblem}.
In summary, our contributions are:
\begin{enumerate}
\item \probtest, a method for unit testing of probabilistic programs with statistical guarantees (\cref{chapter:properties_of_PPs}). 
\item We establish the correctness of the method for testing probabilistic programs (\cref{chapter:properties_of_PPs}), and study its scalability and sensitivity to errors.
\item An implementation of the method as a plug-in for the PyTest framework that can be used for unit testing of probabilistic programs in Python.
\item 
  An empirical evaluation of the effectiveness of \probtest on reinforcement learning applications and a randomized data structure (\cref{chapter:evaluation}). 
  The evaluation demonstrates the effectiveness of \probtest to discover bugs in probabilistic programs used in large systems and practical applications.
\end{enumerate}
\looseness -1

The source code of \probtest, the PyTest plugin and experiments are available in the accompanying artifact~\cite{probtest_artifact}.


\section{Background} \label{chapter:background}

\subsection{Probabilistic Programs}
\label{sec:probabilistic_programs}

Probabilistic programs are programs with stochastic behaviour in their execution. 
That is, programs can make a random choice at any point in their execution; which may result in producing random outcomes for the same input.
In this paper, we use the following formal definition for probabilistic programs:

\begin{definition}\label{def:probabilistic_programs}
  Given an input $i\in I$, a probabilistic program $f$ is described by the probability space $(\Omega_i, \mathcal{F}_i, P_i)$, where
  $\Omega_i$ is the set of all outcomes of the program for input $i$, 
  $\mathcal{F}_i$ is a $\sigma$-algebra such that $\mathcal{F}_i \subseteq 2^{\Omega_i}$ and
  $P_i \colon \mathcal{F}_i \to [0,1]$ is a probability measure.
  We use $\mathcal{O}_i$ to denote the set of outcomes/outputs of the program (the terms outcome and output are used interchangeably throughout the paper).
  Finally, we use $O_i: \Omega_i \to \mathcal{O}_i$ to denote a random variable for the observable outcomes of the program.
  A probabilistic program thus induces a probability distribution over the outcomes of the program for a given input $i$.
  When no confusion arises, we will denote  $(\Omega_i, \mathcal{F}_i, P_i)$ as $(\Omega, \mathcal{F}, P)$ and $O_i$ as $O$.
\end{definition}

The definition above views the execution of the program as a random experiment; i.e., sampling from $O_i$.
Our definition is a simplification of the standard pushforward measure over program statements~\cite{FoundationsOfPP} where we only consider the distribution of the program output for a given input.
Focusing on the program output distribution is sufficient, given our back-box unit testing setup.
We only consider positively almost-surely terminating programs, i.e., programs that terminate with probability 1 and finite expected runtime~\cite{FoundationsOfPP}.
Requiring termination in black-box unit testing is a common restriction, as assertions are checked after the execution of the program.
Note that the probability space of the program only depends on \(i \in I\).
Thus, in a sequence of samples drawn from \(O_i\), each sample is independent and identically distributed.
We will use $f(i)$ to denote the probabilistic program $f$ for an input $i\in I$. 
Note that $f(i)$ is not a function.
Furthermore, the set of inputs $I$ is an abstraction. 
It can represent any combination of input parameters to a program.

\begin{example}
Consider the program in~\cref{lst:bounded_random_walk}, which implements a one dimensional discrete random walk. The program simulates taking $n$ random steps, each step moving either left or right based on a coin toss.
For simplicity, let us consider a random walk consisting of only two steps, i.e., \lstinline|n=2|.
The set of possible outcomes is $\mathcal{O} = \{-2, \ldots, 2\}$ and the $\sigma$-algebra $\mathcal{F}$ is the powerset of the set of outcomes $\mathcal{F} = 2^{\mathcal{O}}$. 
The output random variable $O$ is a distribution defined as $P(O=-2) = P(O=2) = 1/4$ (as $\pm 2$ are only reachable after obtaining two consecutive 0s or 1s from \lstinline|X.rvs()|), $P(O=0) = 1/2$ (as $0$ can be reached by obtaining first 0 and then 1 or first 1 and then 0 from \lstinline|X.rvs()|), and $P(O = 1) = P(O = -1) = 0$ (as $\pm 1$ are not possible outcomes after two steps).
\end{example}

\subsection{Coupon Collector's Problem}\label{subchapter:coverage_research_project}

The coupon collector's problem is a well-known combinatorial problem~\cite{ccp_generalisedCouponCollectorProblem}.
It concerns the problem of drawing balls with replacement from an urn with a finite collection of $N$ different balls until each ball is drawn at least once.
The balls are drawn with probability $p_i\in (0,1]$ with $\sum_{i=1} ^N p_i \leq 1$. When \(\sum_{i=1} ^N p_i \not= 1\), there are one or multiple balls in the urn that are not of interest to the collection. 

Consider the random variable $T$ for the number of draws with replacement to obtain a full collection of the $N$ different balls. 
As $T$ depends on the probability distribution of the balls, we denote it as $T(\textbf{p})$ where $\textbf{p}:=(p_1,\ldots,p_N)$ with $p_i$ representing probability of drawing ball $i$.
The problem is determining the number of balls $k \in \mathbb{N}$ that need to be drawn to obtain a full collection with high probability.
Formally, the problem is formulated as $P(T(\textbf{p})>k)$; i.e., the tail distribution of finding all balls after $k$ trials.
The following theorem establishes an exact solution for the tail distribution~(e.g., \cite{CCP_bounds}).
\begin{theorem}\label{theorem:exact_tail_dist}
Let $k>0$, then $P(T(\textbf{p}) > k)$ is given by
    \begin{equation}\label{eq:ccp_exact_tail_dist}
      P(T(\textbf{p})>k) = \sum_{i=1}^N (-1)^{i+1} \sum_{J: |J|=i} (1-P_J)^k  
    \end{equation}
where $P_J := \sum_J p_i$ and $J$ is a subset of $\{1,...,N\}$. That is, the inner sum is a sum over all elements of size $i$ of the power set of $\{1,...,N\}$. 
\end{theorem}
We map the problem of testing probabilistic programs to the coupon collector's problem and use its theoretical results to develop our unit testing method.


\section{Unit Testing of Probabilistic Programs}\label{chapter:properties_of_PPs}

Here we introduce our unit testing method for probabilistic programs.
We consider unit tests that assert a Boolean property after the execution of the program under test for a given input.
These are standard in unit testing.
An assertion is a map from program outcomes to Boolean values $Q \colon \mathcal{O} \to \{0,1\}$ where $0$ and $1$ denote false and true, respectively.
Composing the assertion with the random variable for the outcome of program under test $Q \circ O$ results in a new random variable over Boolean values; as the assertion is a measurable map~\cite{Measure_Theory}.

The stochastic nature of probabilistic programs requires a probabilistic notion for passing or failing a unit test.
To this end, we introduce the notion of \emph{assertion coverage} for probabilistic programs.
Intuitively, we require that the unit test is executed enough times so that program outcomes satisfying/violating the assertion are induced during executions with high probability.
This ensures that, with high probability, one execution of the unit test will produce an outcome violating the assertion, if there exists such outcome (e.g., in case there is a bug in the program).
The following definition formalizes this notion.

\begin{definition}[Assertion Coverage for Probabilistic Programs] \label{def:coverage_of_PPs}
  Let $f(i)$ denote a probabilistic program under test $f$ for input $i \in I$ described by the probability space $(\Omega,\mathcal{F},P)$.
  Let $O \colon \Omega \to \mathcal{O}$ be the random variable for $f(i)$.
  Let $Q \colon \mathcal{O} \to \{0, 1\}$ be a measurable map corresponding to an assertion (i.e., a Boolean property on program outcomes) for the program under test.
  Let $T$ be a random variable for the number of times to run the program to see all assertion outcomes, i.e., $\{0, 1\}$.
  Given $\epsilon>0$ and $k>0$, we say that a unit test achieves assertion coverage after executing the program $k$ times with probability $1-\epsilon$ iff 
  \begin{equation}
    P(T>k)<\epsilon.
  \end{equation}
  We refer to $P(T>k)$ as the tail distribution of $T$.
\end{definition}

In what follows, we present, \probtest, our black-box unit testing method for probabilistic programs, and we prove that it satisfies assertion coverage.

\Cref{alg:probtest} shows the pseudo-code of our method to assert Boolean properties over outcomes of probabilistic programs.
The algorithm takes four inputs.
The first two inputs are: the probabilistic program under test $f(i)$ (for program input $i \in I$) and the assertion to test $Q \colon \mathcal{O} \to \{0,1\}$.
These inputs are standard in unit testing.
Additionally, the method takes as input an \emph{assertion specification} vector $\textbf{p} = (p, 1-p)$, where $p \in (0,1)$ is the \emph{assertion violation probability}.
This parameter specifies the maximum allowed probability for the assertion to be violated.
For instance, a requirement for the program can be that \emph{``property $Q$ must hold with probability 0.99''}.
In this case, the assertion violation probability must be set to $p = 0.01$, and, consequently, $\textbf{p} = (0.01, 0.99)$.
The last input is the error probability $\epsilon > 0$, which indicates the accuracy of the test finding a violation if one exists.
Lower values of error probability result in higher accuracy.

The last two input parameters $p$ and $\epsilon$, result from the stochasticity of the probabilistic program.
Although these parameters are not standard in regular unit testing, they commonly appear in the analysis of probabilistic systems (cf.~\cref{chapter:related_work}).
We remark that, even though $p$ and $\epsilon$ cannot be set to 0, they can be as small as needed.
This is a decision that the tester takes based on the requirements for the program under test.
In~\cref{sec:scalability_k}, we study the scalability and accuracy trade-offs of selecting different values for these parameters.

\begin{algorithm}[t]
  \hspace*{\algorithmicindent}\textbf{Input:}~Input value $i \in I$; Program under test $f(i)$ described by $(\Omega_i, \mathcal{F}_i, P_i)$ \\
  \hspace*{\algorithmicindent}\phantom{\textbf{Input:}}~Assertion $Q \colon \mathcal{O} \to \{0,1\}$; Assertion spec. $\textbf{p} = (p, 1-p)$ with $p \in [0,1]$; and \\
  \hspace*{\algorithmicindent}\phantom{\textbf{Input:}}~Error probability $\epsilon > 0$  \\
  \hspace*{\algorithmicindent}\textbf{Output:}~Unit test result $r \in \{0,1\}$; pass ($1$) or fail ($0$)
  \begin{algorithmic}[1]
    \Procedure{Probtest}{$i, f(i), Q, \textbf{p}, \epsilon$}
      \State $k \gets \arg \min_{k \in \mathbb{N}} P(T(\vec{p}) > k) < \epsilon$ \Comment \textit{To be determined using Theorem~\ref{eq:ccp_exact_tail_dist}}\label{line:ccp}
      \For{$1$ to $k$} \label{line:start_repeated_runs}
        \State $o \; \sample \; f(i)$ \Comment \textit{The symbol \(\sample\) denotes sampling an output from \(f(i)\)} \label{line:execute_program}
        \If{$\neg Q(o)$}
          \State \Return $0$ \label{line:assertion_not_satisfied}
        \EndIf
      \EndFor \label{line:end_repeated_runs}
      \State \Return $1$ \label{line:assertion_satisfied}
    \EndProcedure
  \end{algorithmic}
  \caption{Black-box unit testing method pseudo-code.} \label{alg:probtest}
\end{algorithm}

\Cref{alg:probtest} uses Theorem~\ref{eq:ccp_exact_tail_dist} to automatically determine the number of times ($k$) that the unit test must be executed to achieve assertion coverage with probability less than $1-\epsilon$ (line~\ref{line:ccp}).
In our implementation, we incrementally explore values of \(k\) starting from \(k=2\) until \( P(T(\vec{p}) > k) < \epsilon \) holds. 
This naive search exhibits good practical performance. 
For instance, iterating up to values of \(k\) in the order of \(10^7\) takes less than 2 seconds.
However, more efficient methods to compute \(\arg \min\) can be used for this step of the algorithm.
In lines \ref{line:start_repeated_runs}-\ref{line:end_repeated_runs} we run the probabilistic program up to $k$ times.
Line~\ref{line:execute_program} denotes executing the probabilistic program.
If the output does not satisfy the assertion, we stop the execution of the unit test and return false (line~\ref{line:assertion_not_satisfied}); as we have found an error in the program.
If after $k$ executions we find no program output violating the assertion, we return true (line~\ref{line:assertion_satisfied}).
This indicates that the assertion is satisfied (with probability $1-\epsilon$).

\begin{example}
  Consider again the \texttt{n\_random\_walk} program in \cref{lst:bounded_random_walk}, and the property stating that the value of \texttt{pos} does not exceed a boundary $b$ with certain probability.
  Here we discuss how to apply \probtest to test this property.
  
  In this example, $f(i)$ is the probabilistic program \texttt{n\_random\_walk} and the program input, \lstinline|n=20|, determines the number of steps of the random walk.
  The assertion $Q$ is a Boolean predicate checking whether the position after executing the 20 steps is below some boundary $b$.
  We arbitrarily use \(b = 10\) for this example.
  The following listing shows the code for the complete unit test:
   \vspace{-3mm}
  \begin{lstlisting}[
    language=python]
    def test(n: int = 20, b: int = 10):
        o = n_random_walk(n) # f(i)
        assert o < b # Q
  \end{lstlisting}
  To determine how many times this test must be executed, we define the assertion specification $\textbf{p}$ and probability of error $\epsilon$.
  We illustrate two possibilities for a tester to decide on the values for these parameters: based on system requirements or based on expected program behavior.
  
  For the system requirements case, assume high level system requirements that specify the required robustness of the system.
  For example, \emph{``with probability 0.99 the random walk must not finish in a position more than 10 steps away from the origin.''}
  Probabilistic requirements on the length of a random walk are commonplace in MCMC algorithms for Bayesian inference~\cite{brooks2011handbook}.
  To check this requirement, the tester must set $\textbf{p} = (0.01, 0.99)$.
  We use a value of $\epsilon = 0.02$, which defines the accuracy of the method.
  This value of $\epsilon$ means that if the probability for the random walk to finish 10 steps away from the origin is more than $0.01$, then a violation of the property will be triggered with probability $0.98$ (i.e., $1-\epsilon$) if one exists.
  A tester may choose arbitrarily lower values of $\epsilon$ (as long as they are larger than 0) to increase accuracy; as expected, lower $\epsilon$ values result in larger number of executions of the test (cf.~\cref{sec:too_large_p}).
  For $\textbf{p} = (0.01, 0.99)$ and $\epsilon=0.02$, our algorithm requires to execute the test $k=390$ times.

  The tester can also select $\textbf{p}$ and $\epsilon$ based on their knowledge about the program.
  For example, the documentation may state that the random walk moves to adjacent states with probability 1/2.
  Thus, it is a possible to finish in a position greater than 10 with probability $1/2^{10}$.
  To determine whether this is a possible behavior in the program, the tester sets $\textbf{p}=(1/2^{10}, 1-1/2^{10})$ and the same accuracy as before \(\epsilon = 0.02\)
  In this scenario, the test must be executed $k=4004$.
  Note that reducing the probability of violating the assertion (from $0.01$ to $1/2^{10}$) requires increasing the number of executions (cf.~\cref{sec:too_large_p}).
  As before, this number of executions guarantees that if the assertion can be violated with probability $1/2^{10}$, then with probability $0.98$ we will observe a program outcome violating the property in one of the $k=4004$ executions. \qed

\end{example}

The correctness of \probtest follows from the coupon collector problem.

\begin{theorem} \label{theorem:correctness_of_method_coverage}
  Let $f(i)$ be a probabilistic program with input $i \in I$, $\textbf{p}$ be an assertion specification, and $\epsilon > 0$ the probability of error.
  Let $k \in \mathbb{N}$ be the number of times \cref{alg:probtest} executes the test (defined in line~\ref{line:ccp}), and $T(\textbf{p})$ be the random variable for the required number of program (test) executions to observe all outcomes of $f(i)$.
  Then, $P(T(\textbf{p}) > k) < \epsilon$, i.e.,~\cref{alg:probtest} achieves test assertion coverage for $f(i)$ with probability $1-\epsilon$.
\end{theorem}

\begin{proof*}
  The theorem follows by mapping the elements of the unit testing method in~\cref{alg:probtest} to those of the coupon collector's problem.
  The set of balls maps to the set of property outcomes ($\{0,1\}$), the distribution for drawing each ball maps to the distribution of property outcomes ($\textbf{p}$), and each draw (with replacement) maps to an execution of the unit test ($k$).
  Since~\cref{alg:probtest} uses Theorem~\ref{theorem:exact_tail_dist} to determine a value of $k$ such that $P(T(\textbf{p}) > k) < \epsilon$ and it runs the test $k$ times, then it directly follows that test assertion coverage (\cref{def:coverage_of_PPs}) holds with probability $1-\epsilon$, as required.
\end{proof*}

\begin{figure}[t!]
  \centering
  \includegraphics[width=0.5\textwidth, trim={3mm 2mm 2mm 2mm}, clip]{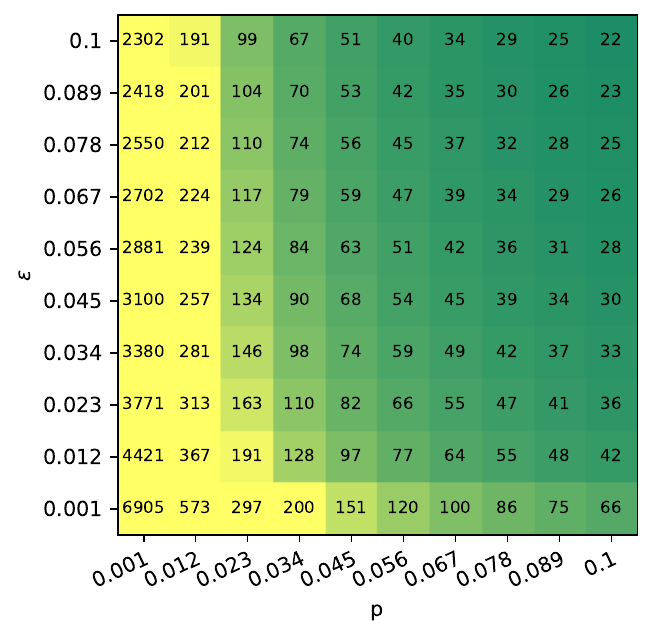}
  \caption{Values of $k$ for \(p, \epsilon \in (0.001, 0.1)\)}
  \label{fig:scalability_of_method}
\end{figure}

\subsection{Scalability of Test Executions} \label{sec:scalability_k}
Computational resources may impact the assertion violation probability $p$ or error probability $\epsilon$ values that can be used in practice.
Although the value of $k$ can be computed efficiently using~\cref{theorem:exact_tail_dist} (since $N$ is the main factor determining the complexity of computing $k$ and we are in the case for $N=2$), the value of $k$ increases rapidly as the values of $p$ and $\epsilon$ decrease; especially for low \(p, \epsilon\) values.
In other words, as the tester sets higher requirements for the correctness of the assertion, the method returns a increasingly larger $k$.
\Cref{fig:scalability_of_method} shows the rapid growth of $k$ for $p, \epsilon \in (0.001, 0.1)$.
We observe, for instance, that if a test can be executed at most 60 times then the accuracy (\(1-\epsilon\)) can be at most \(97.6\%\) and the assertion violation probability \(0.056\).
However, executing unit tests thousands of times is commonplace, and the plot shows that with less than \(7000\) test executions we achieve \(99.9\%\) accuracy for an assertion violation probability of \(0.001\).
In~\cref{chapter:evaluation} we demonstrate that \probtest\ can be effectively used in realistic applications.

\subsection{Sensitivity to Error in Assertion Violation Probabilities} \label{sec:too_large_p}
Let $\hat{p}$ denote an assertion violation probability, for an assertion $Q$, that differs from the real assertion violation probability $p$ given by the program implementation.
In this situation, a natural question is: \emph{What is the probability that our method triggers an outcome violating the assertion if $\hat{p}$ is used instead of \(p\) in~\cref{alg:probtest}?}
The answer depends on whether \(\hat{p}\) under- or overestimates $p$.

If a tester underestimates the value of $p$, i.e., $\hat{p} < p$, then, for any $\epsilon$, the value of $k$ may be larger than necessary.
This is due to the monotonic increase of \(k\) for decreasing values of \(\epsilon > 0\) and \(p \in (0,0.5]\).
This trend is symmetric for \(p > 0.5\) as \(1-p < 0.5\) (cf.~\cref{theorem:exact_tail_dist} for \(N=2\)).
Thus, our method will require a value $\hat{k}$ such that $\hat{k} \geq k$ (where $\hat{k}$ and $k$ correspond to $\hat{p}$ and $p$, respectively).
But the impact of using $\hat{k}$ is minimal.
\Cref{alg:probtest} finishes after an assertion violation is found, and theorems~\ref{eq:ccp_exact_tail_dist} and~\ref{theorem:correctness_of_method_coverage} show that this must happen in at most $k$ executions.
Therefore, our algorithm will repeat the test only $k$ times.
The only overhead of using $\hat{p}$ is that it takes longer to find $\hat{k}$, as it is a larger value than $k$.
Since determining $\hat{k}$ can be done efficiently, this overhead is minimal.

If the tester overestimates the assertion violation probability, i.e., $\hat{p} > p$, then the value of $k$ will be smaller than necessary; again due to the monotonic increase of \(k\) for decreasing \(p, \epsilon\).
This implies that the probability of executing a violating outcome will be smaller than if $p$ had been used instead of $\hat{p}$.
This is problematic, as it detriments the chances of the tester finding errors in the program under test.
\looseness -1

\begin{figure}[t!]
  \centering
  \includegraphics[width=.51\textwidth, trim = 5mm    1mm 15mm 15mm, clip]{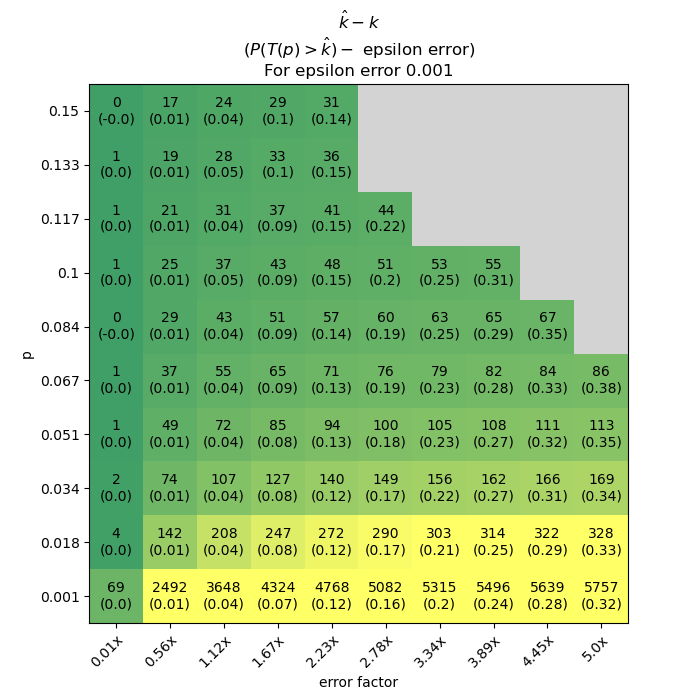}
  \includegraphics[width=.46\textwidth,   trim = 22.5mm 1mm 15mm 15mm, clip]{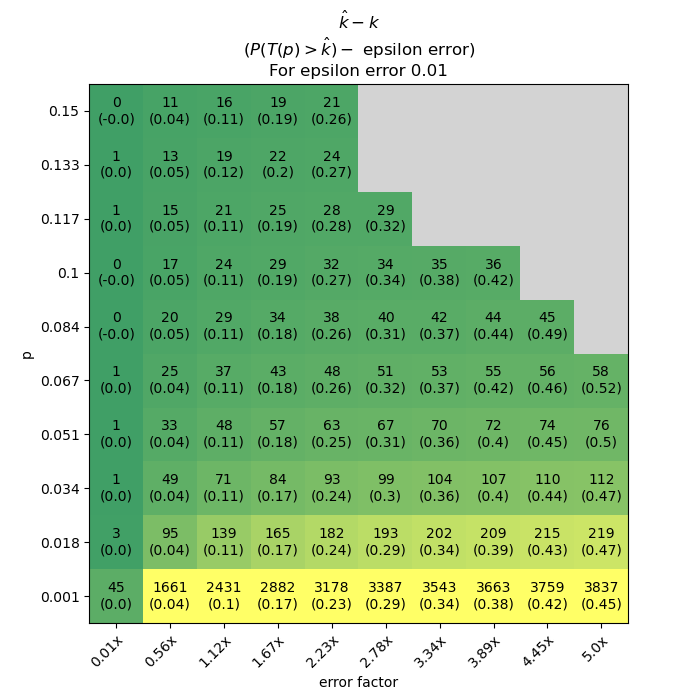}
  \caption{
    Effect of overestimating $p$ on the value of $k$ and the probability of finding a violating outcome.
    The y-axis is the real assertion violation probability $p$.
    The x-axis displays the factor of error, $\phi$, for the overestimated assertion violation probability, $\hat{p} = p + p \cdot \phi$.
    Each cell has two rows: the upper row shows $k - \hat{k}$ (corresponding to $p$ and $\hat{p}$, respectively), and the bottom row shows the distance between the tail probability for $\hat{k}$ the error probability, $P(T(\textbf{p}) > \hat{k}) - \epsilon$.
    We explore 2 error probabilities $\epsilon \in \{0.001, 0.01\}$.
    Gray cells corresponds to cases with $\hat{p} > 0.5$.
    These are symmetric to $1-\hat{p}$ and are already covered in the plot, as $1-\hat{p} < 0.5$.
    \looseness -1
  }
  \label{fig:overestimating_p}
\end{figure}

In~\cref{fig:overestimating_p}, we explore the effect of overestimating $p$ on the probability of executing a violating outcome.
We consider the effect of overestimating $p$ by a factor $\phi$ (i.e., $\hat{p} = p + p \cdot \phi$), for different values of $p$ and for different error probabilities $\epsilon$.
For each case (cells in the figure) we show the difference in number of executions $k - \hat{k}$ and the difference between the tail probability and the specified probability of error $P(T(\textbf{p})>\hat{k}) - \epsilon$.
The latter quantifies the decrease in the probability of executing an outcome violating the assertion if $\hat{p}$ is used instead of $p$.
We consider small values $p$ and $\epsilon$, as this is where more variation occurs (cf.~\cref{sec:scalability_k}).
\looseness -1

For a very small overestimation $\phi = 0.01$ (first column in~\cref{fig:overestimating_p}), we observe very small difference in number of executions ($k - \hat{k}$)  for all $\epsilon \in \{0.001, 0.01\}$, with only notable changes in the case $p=0.001$.
Despite this difference, note that the difference between tail probability and probability of error is $0$ for all cases.
This means that overestimating $p$ by a factor of $\phi = 0.01$ has no impact on the probability of executing a violating outcome.

We observe in~\cref{fig:overestimating_p} that the distance between number of executions $k - \hat{k}$ heavily depends on the value of $p$ for all $\epsilon$ values.
Lower values of $p$ result in higher distance between number of executions as $\phi$ increases.
This is expected due to the growth in the value of $k$ for small values of $p$ and $\epsilon$ that we observed in~\cref{fig:scalability_of_method}.
However, the large difference between number of executions does not necessarily imply a large decrease in executing a violating outcome.

Interestingly,~\cref{fig:overestimating_p} shows that the distance between the tail probability and $\epsilon$ is very similar (and often the same) for all columns.
This indicates that changes on the probability of executing a violating outcome are mainly affected by the error factor $\phi$, and not by the value of $p$.
Nevertheless, the value of $\epsilon$ affects the magnitude of the distance between the tail probability and $\epsilon$.
For instance, in order for the distance to be $\approx0.10$ (i.e., it is around $10\%$ less likely to execute the outcome violating the assertion), the error factor can be up to $\phi \leq 1.67$ for $\epsilon = 0.001$ and up to $\phi \leq 1.12$ for $\epsilon=0.01$.
This effect generalizes to any distance, i.e., using smaller $\epsilon$ values always produces smaller decrease on the probability of executing a violating outcome for the same error factor $\phi$. 
This insight reveals that, if testers are unsure about the value of $p$, it is recommended to select an $\epsilon$ value that is as small as possible; as this ensures a higher probability of finding errors in case $\hat{p}$ is underestimating $p$.

\subsection{Implementation} \label{chapter:implementation}

We have implemented \emph{ProbTest} as a plugin for the Python testing framework \textit{Pytest}~\cite{probtest_artifact,pytest}.
To apply the plugin, the tester must provide:
\begin{inparaenum}[i)]
    \item the system under test as a Python program;
    \item a test suite for the system under test for an input written in \textit{Pytest} (the test suite may consist of a finite number of tests, each testing a property of the system);
    \item an assertion specification $\textbf{p}=(p,1-p)$ where $p \in (0,1)$ is the assertion violation probability; and
  \item the error probability $\epsilon > 0$.
\end{inparaenum}
The plugin automatically determines the number of times to execute each test in the test suite and runs the tests sequentially.
The execution finishes as soon as one test fails; as described in~\cref{alg:probtest}.
The source code for the \probtest plugin is available in the accompanying artifact.


\section{Case Studies} \label{chapter:evaluation}

In this section, we demonstrate using \probtest in two realistic case studies: a randomized data structure (\cref{chapter:skip_list}) and a reinforcement learning system (\cref{chapter:frozen_lake}).
The accompanying artifact contains the source code for all experiments.


\subsection{Skip List} \label{chapter:skip_list}

\begin{wrapfigure}{r}{.25\textwidth}
\vspace*{-8mm}
  \begin{tikzpicture}[
    auto,
    start chain,
    every node/.style={font=\small},
    item/.style={rectangle,minimum height=4mm,minimum width=4mm,thick,draw=black},
    label/.style={rectangle,minimum size=6mm},
    every join/.style={->}
    ]
    \matrix[row sep=2mm, column sep=2mm]{
      \snode{2head}{\bullet}; & & \snode{2node2}{2}; & \snode{2end}{\texttt{None}};\\
      \snode{1head}{\bullet}; & & \snode{1node2}{2}; & \snode{1end}{\texttt{None}};\\
      \snode{0head}{\bullet}; &\snode{0node1}{1};  & \snode{0node2}{2}; & \snode{0end}{\texttt{None}};\\
    };

    {
      [start chain] \chainin(0head); \chainin(0node1) [join]; \chainin(0node2) [join]; \chainin(0end) [join];         
      [start chain] \chainin(1head);  \chainin(1node2) [join]; \chainin(1end) [join];
      [start chain] \chainin(2head); \chainin(2node2) [join]; \chainin(2end) [join];

    }
    {
      [start chain] \chainin(2node2); \chainin(1node2) [join]; \chainin(0node2) [join];
    }
  \end{tikzpicture}
  \caption{Skip list with 2 nodes and 3 levels.}
  \label{fig:skip_list_example}
\vspace*{-5mm}
\end{wrapfigure}
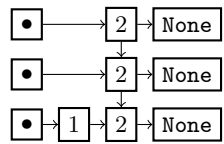


A \emph{skip list} is a randomized data structure consisting of a finite collection of ordered linked lists of nodes \(L_1, \ldots, L_M\)~\cite{RandomizedAlgorithms}.
We write \(n \in L_i\) to denote list \(L_i\) contains node \(n\).
Each \(L_i\) denotes a \emph{level}.
Each level contains a subset of the nodes from the level below, i.e., \(L_M \subseteq L_{M-1} \subseteq \ldots \subseteq L_1\).
Thus, the first level \(L_1\) contains all nodes in the skip list, and the size of the skip list is defined as \(R=|L_1|\).
A node in \(L_i\) has a pointer (\texttt{next}) to the next node in the link list, and, if \(i>1\), a pointer (\texttt{lower}) to a node with the same key in the level below \(L_{i-1}\).
The level of a node \(n\) is defined as \(\max(\{i \mid n \in L_i\})\).
\Cref{fig:skip_list_example} depicts an example skip list with 2 nodes and levels \(L_1, L_2, L_3\). 
Node 1 has level 1 and node 2 level 3.
The symbol \(\bullet\) denotes the head of each \(L_i\) and \texttt{None} is the \texttt{next} value for nodes at the tail.
One of the benefits of skip lists is the expected time \(O(\log(R))\) for insert, delete and search~\cite{RandomizedAlgorithms}.
Various industrial applications use skip lists~\cite{skip_list_application_robot_navigation,skip_list_application_discord,skip_list_application_MemSQL}.
Due to the randomized nature of skip lists, standard unit testing may be ineffective.
We use \probtest to test a skip list implementation (with statistical guarantees), and demonstrate its applicability for testing software used in industry.

The randomized behavior of skip lists originates in insertions.
When inserting a node, the node is first added to \(L_1\) in its correct place according to the order among nodes.
Then, a coin with bias \(p_r\) is tossed.
If the outcome is heads, the node is added to the layer above (also preserving order).
Otherwise, the insertion finishes.
The coin toss is repeated until the outcome is tails.
Consequently, executing an insertion produces a set possible skip lists; depending on its randomized behavior.
\looseness -1

To test our skip list implementation, we have designed a battery of 26 unit tests.
The test are split into 4 categories: validity, post-condition, metamorphic and equivalence.
Validity tests check that after executing an insertion, deletion or search the skip list remains in a valid state.
According to~\cite{RandomizedAlgorithms}, a skip list is valid if the following properties hold:
\begin{inparaenum}[i)]
\item the maximum number of layers is never exceeded;
\item each layer is a subset of the layer below it;
\item all layers are ordered;
\item all inserted elements should be in the bottom layer;
\item No node in \(L_1\) have pointers to nodes below them; and
\item Nodes in \(L_i\) with \(i>1\) must have a pointer the layer below \(L_{i-1}\) if it contains a node with the same key.
\end{inparaenum}
Post-condition tests check that operations behave as  expected.
For instance, after deleting a node, a search for said node returns \texttt{None}.
Metamorphic properties encode properties between two lists that must hold before and after executing a set of operations.
For example, consider two lists \(l_1, l_2\) that contain the same nodes.
One of our metamorphic tests checks that after adding a new node to \(l_1\) and \(l_2\) the lists still contain the same elements---although they may be structurally different due to randomness.
Finally, equivalence properties are similar to metamorphic properties for list equivalence but also include properties checking cases where the resulting lists must not be equivalent.
For instance, consider again \(l_1, l_2\). 
If we add an element only to \(l_1\), then \(l_1\) and \(l_2\) must not contain the same nodes.
\Cref{appendix:skip_list_properties} contains the details of each of the 26 properties.

The next step is to define the assertion violation probability \(p\).
The goal is to select a value of \(p\) ensuring that all possible skip lists resulting from a series of insertions are tested at least once.
To this end, we determine the probability of generating the most unlikely skip list after a series of insertions.
This ensures that we will check the properties above for this unlikely skip list (with high probability).
Also, since any other resulting skip list have higher probability, they will be covered by at least one execution (cf.~\cref{sec:too_large_p}).

The probability of generating the most unlikely skip list depends on the number of possible skip lists and the probability of raising a node.
Since each node can be raised up to $M$ levels, then there are $M^R$ possible skip lists after $R$ insertions.
Let \(p_r\) denote the probability of raising a node.
When $p_r \leq 1/2$, the least likely state is when all $R$ nodes are raised to the maximum layer. 
When $p_r > 1/2$, however, the least likely state is when all nodes are raised to the layer just below the maximum layer, $L_{M-1}$. 
This corresponds to raising the node repeatedly $M-2$ times with probability $p^{(M-2)}$, followed by a final coin toss with the result of not raising the node with probability $(1-p)$. 
In summary, we define the assertion violation probability as
\begin{equation}
p =\begin{cases}
          p_r^{R(M-1)} \quad &\text{if } p_r \leq 1/2 \\
          p_r^{R(M-2)}(1-p_r)^R \quad &\text{otherwise.} \\
     \end{cases}
\label{eq:min_skiplist_spec}
\end{equation}

To evaluate the effectiveness of \probtest, we injected 8 bugs in our skip list implementation.
The bugs are designed to appear for an increasing number of insertions \(R = 1 .. 4\).
There are two bugs for each \(R\).
The goal is to evaluate the ability of \probtest to find bugs that appear with decreasing probability.
Each bug is unique, and they may be triggered by one or many of the skip lists that result from running a test.
For instance, one of the injected bugs fails to delete nodes at the top layer.
This bug occurs only in one of the possible skip lists that result after 1 insertion.
Another injected bug incorrectly raises the level of a set of nodes to \(L_{i+1}\) if there are 3 or more nodes at level \(L_i\).
In this case, the bug may be triggered by more than one of the resulting skip lists resulting from 3 insertions.
We refer interested readers to~\cref{appendix:skiplist_bugs} for details of each bug.

\begin{figure}[t!]
\centering
\scalebox{1.0}{
\begin{tabular}{
  c@{\hskip 2mm}
  c@{\hskip 2mm}
  c@{\hskip 1mm}
  c@{\hskip 0mm}
}
\toprule
bug & \(R\) & \(k\) & Prob. find bug \\
\midrule
1 & 1 & 11  & 1.000 \\
2 & 1 & 11  & 0.997 \\
3 & 2 & 47  & 0.955 \\
4 & 2 & 47  & 1.000 \\
5 & 3 & 191 & 0.997 \\
6 & 3 & 191 & 1.000 \\
7 & 4 & 767 & 0.960 \\
8 & 4 & 767 & 1.000 \\
\bottomrule
\end{tabular}
}
\caption{Estimated probability of finding injected bugs with \probtest (1000 executions).}
\label{tbl:probability_finding_skip_list_bugs}
\end{figure}


\paragraph{Experiments and results.}
We consider a skip list with node raising probability \(p_r = 1/2\) and at most \(M=3\) levels, the battery of 26 unit tests and the 8 injected bugs.
Given this setup and~\cref{eq:min_skiplist_spec}, the assertion violation probability (\(p\)) equals \(0.25, 0.0625, 0.0156, 0.0039\) for $R$ equals 1 to 4, respectively. 
We consider a probability of error \(\epsilon = 0.05\).
We run each experiment 1000 times and estimate the probability that \probtest finds the injected bugs.
As a sanity check, we first run the experiments with no injected bugs for 1-4 insertions.
As expected, \probtest finds no bugs.
\Cref{tbl:probability_finding_skip_list_bugs} shows the results for injected bugs.
The first two columns show bug ID and the number of insertions required to trigger the bug.
The third column shows the number of times that \probtest executes the test.
The last column shows the estimated probability of \probtest finding the bug.
We can observe that, for most bugs, \probtest finds the bugs with probability \(\geq 0.99\).
Only for bugs 3 and 7 the probability decreases to \(0.955\) and \(0.96\), respectively.
These bugs corresponds to those that are only triggered with very low probability.
Nevertheless, all bugs are detected with probability at least \(0.95\), i.e., \(1-\epsilon\).
These results are a consequence of \cref{theorem:correctness_of_method_coverage} and selecting a value of \(p\) that is lower or equal to the probability of generating the skip list with lowest probability after \(R\) insertions.

All in all, we demonstrated the use of \probtest to effectively test a realistic randomized data structure that is used in industrial applications, namely, the skip list. 
Furthermore, our results show that \probtest finds bugs with the desired statistical guarantees.
\looseness -1


\subsection{Reinforcement Learning Applications} \label{chapter:frozen_lake}

\begin{table}[t!]
\centering
\scalebox{1.0}{
\begin{tabular}{P{2.5cm} l c P{1.6cm} P{2.2cm} P{2.2cm}}
\toprule
 & \(p\) & Property & \makecell{Num.\\episodes} & \makecell{Prob. property\\does not hold} & \makecell{Prob. Probtest\\finds bug} \\
\midrule
frozen lake 4x4 & 0.01 & Q1 & 100 & 0.000 & 0.000 \\
frozen lake 4x4 & 0.001 & Q2 & 100 & 0.002 & 0.990 \\
frozen lake 7x7 & 0.01 & Q1 & 25000 & 0.000 & 0.000 \\
frozen lake 7x7 & 0.001 & Q2 & 25000 & 0.060 & 1.000 \\
cliff walking & 0.001 & Q & 25000 & 0.007 & 1.000 \\
\bottomrule
\end{tabular}
}

\caption{Summary of RL experiment results. Left to right: assertion violation probability in specification (\(p\)), number of training episodes, estimated probability that the property fails over 100k runs, probability that \probtest finds failure over 100 runs.}
\label{tbl:rl_experiment_results}
\end{table}

We consider two RL case studies from the Gymnasium Python library \cite{towers2024gymnasium}; one of the most widely used libraries to develop RL agents in Python. Gymnasium environments serve as
testbeds for RL algorithms. 

\paragraph{Frozen lake.} The environment involves an $n\times m$ board that represents a frozen lake with holes distributed on the lake. An agent can move around the board until it falls into a hole or reaches a goal state. The holes and goal states are terminal states. We consider two board sizes 4x4 and 7x7 provided in the Gymnasium library documentation. The board has stochastic behavior representing slippery ice, which means the agent moves in the intended direction with a probability of 1/3, and (slipping) in one of the two perpendicular directions with a combined probability of 2/3.

\paragraph{Cliff walking.} This is another classic RL example in which an agent must navigate a map while avoiding falling off a cliff that spans the edge of the map. If the agent steps into the cliff area, it receives a negative reward and is sent back to the initial state. The environment is episodic and includes stochastic dynamics.

Reinforcement learning agents learn a policy to achieve a goal by interacting with an environment using RL algorithms. This process involves training the agent through repeated episodes, where it explores actions and updates its behavior based on feedback from the environment. We use Q-learning \cite{watkins1992q}, a traditional RL algorithm, which updates a Q-table to approximate the optimal state-action-values (find the best action) based on observed transitions and rewards during training. The algorithm implementation uses a standard exploration strategy in which actions are selected at random with a small configurable probability and otherwise the best action is selected according to the policy.
\looseness -1

\paragraph{Experiments and results. }
We are interested in studying whether we can discover errors in the behavior of an agent trained in these stochastic environments.
Due to the randomness in the environment, traditional unit testing methods are not applicable for this type of systems.
Here we demonstrate the use of \probtest to test reinforcement learning applications.
The experiments are designed to assess whether \probtest can be used to test reinforcement learning policies.
For the case studies, a \emph{policy} determines the actions (movements) of agent, and the \emph{environment} determines the state (position) of the agent given a selected action.
Note that source of randomness is in the environment, but it could also be in the policy due to the small probability of selecting a random action as explained above.
Since our method is black-box, the exact source of randomness is irrelevant. 
We do not directly inject bugs in the program, as the lack of enough training is considered to be the source of buggy behavior in the experiments (and in practical applications).
For the frozen lake we test the following properties: \emph{Q1: The agent never falls into holes in $s$ steps following policy $\pi$}, and \emph{Q2: The agent never takes more than $s$ steps before reaching a terminal state following policy $\pi$}. For $s$ we use values 100 and 200 for the 4x4 and 7x7 boards (these values are selected based on the episode lengths recommended by the Gymnasium library documentation). 
To test these properties using \probtest, we must provide an assertion specification for each program property. 
Unlike for the skip list, here we consider that the assertion violation probability is defined by high-level requirements.
This type of probabilistic high-level requirements are  common in practical applications of RL.
Let $P(\neg Q1)$ and $P(\neg Q2)$ denote assertion violation probabilities for $Q1$ and $Q2$, respectively.
We consider $P(\neg Q1)=0.01$ and $P(\neg Q2)=0.001$. 
That is, \emph{``the agent must not fall into holes with probability 0.99''} and \emph{``the agent must reach a terminal state before 100 steps with probability 0.999''}, respectively.
These values have been chosen arbitrarily small on the basis that we expect the agent to be faster at learning to avoid holes than taking too many steps. 
Similarly for the cliff walking, we test the property \emph{Q: The agent survives without falling off the cliff for at least s steps}, with $s=30$ and $P(\neg Q)=0.001$. 
The value of the parameters are selected proportionally to the map size of cliff walking in the Gymnasium library.

\begin{figure}[t]
  \centering
  \includegraphics[width=.49\textwidth]{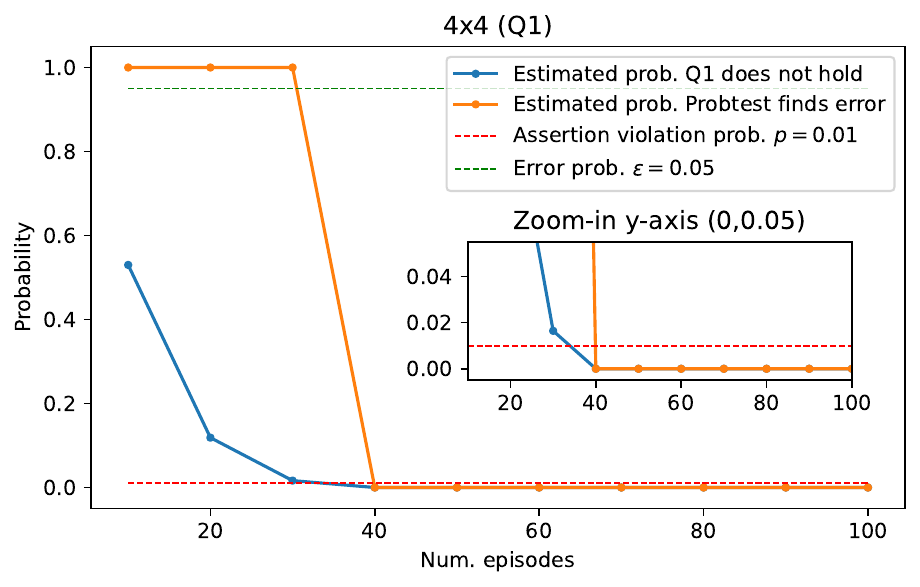}  
  \hfill
  \includegraphics[width=.49\textwidth]{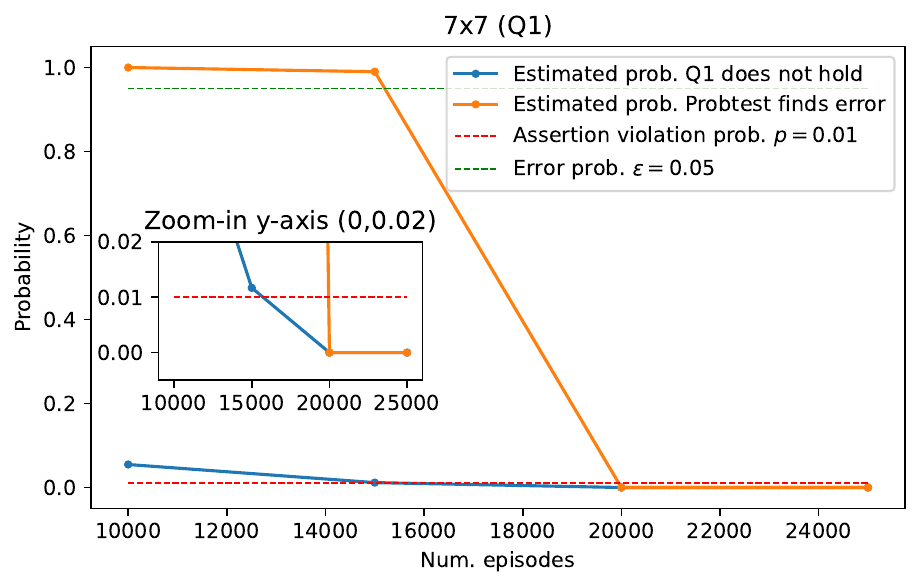}
  \caption{Estimated policy violation probability for Q1 (blue) and probability of finding errors (orange) for the frozen lake problem using policies trained on increasing number of episodes. The red dashed line marks the assertion violation probability \(p\) and the green line \(1-\epsilon\).}
  \label{fig:two_images}
\end{figure}

A summary of the results of experiments is presented in \cref{tbl:rl_experiment_results}. 
First, we train each RL agent for a number of episodes (third column in~\cref{tbl:rl_experiment_results}) selected considering the Gymnasium documentation. 
Then, we estimate the probability of property violations by evaluating the policy for 100k runs (column 4 in~\cref{tbl:rl_experiment_results}).
This value is used to approximate the distance between the real probability of a violation to the value of \(p\) that we selected for \(Q1\)  and \(Q2\).
Finally, we run \probtest using the assertion violation probabilities above and $\epsilon = 0.05$ (as before).
\probtest executes the test 299 times for property $Q1$ and 2995 times for the other properties. 
The last column in~\cref{tbl:rl_experiment_results} shows the estimated probability of \probtest finding a violation (computed over 100 runs).
We observe that \probtest finds violations with very high probability (\(\geq 0.99\)) when the estimated probability of failure is above the assertion violation probability.
This is a notably high accuracy given our \(\epsilon=0.05\), which only requires an accuracy \(\geq 0.95\).
Additional experiments show that results also apply for different numbers of training episodes.
\Cref{fig:two_images} plots the results for increasing number of training episodes for the frozen lake problem (both maps) and property Q1.
As expected, the estimated probability that Q1 does not hold for the learned policy decreases as the number of training episodes increases.
Similar to \cref{tbl:rl_experiment_results}, the figure shows that \probtest detects with very high probability when Q1 can be violated, given the assertion violation probability.
Even in cases where the estimated probability of failure is close to the assertion violation probability (these cases are zoomed-in in the plots), \probtest finds failures with probability significantly larger than \(1-\epsilon = 0.95\).
We refer interested readers to \cref{appendix:more_rl_results} for the complete list of RL results.
All in all, these experiments demonstrate the effectiveness of \probtest to find violations in such RL applications.
\looseness -1


\section{Related Work} \label{chapter:related_work}

We discuss the most relevant works on testing probabilistic programs, but also formal verification techniques focused on probabilistic programs.

We start discussing model-based methods, i.e., methods that require a probabilistic model of the program such as a Markov Decision Process~\cite{MDP}.
Probabilistic model checking is a formal verification technique that has been studied for many years~\cite{DBLP:conf/cav/KwiatkowskaNP11,DBLP:journals/sttt/HenselJKQV22}. 
Probabilistic model checking algorithms exploit algebraic methods to verify probabilistic properties~\cite{DBLP:books/daglib/0020348}.
These methods support expressive property languages such as PCTL~\cite{DBLP:journals/fac/HanssonJ94} where assertion coverage and much more complex properties can be expressed and verified.
Prasetya and Klomp use probabilistic model checking to propose a model-based method for test coverage with statistical guarantees~\cite{pp_coverage_modelbased}. 
The method can assess stricter coverage notions than assertion coverage such as path coverage.
Another line of work is statistical model-checking (SMC) \cite{StatisticalModelChecking}.
Given a probabilistic model, this method uses Monte Carlo simulation and statistical tests to verify whether a property holds with statistical guarantees~\cite{IntroductionToStatisticsAndDataAnalysis,10.1007/978-3-031-90643-5_9,meggendorfer2025oddsimprovingfoundationsstatistical}.
Similar to the above, SMC methods support PCTL and other expressive property languages.

The main difference between the above works and \probtest is the need for a model.
Obtaining models from source code is challenging and often requires introducing abstractions.
As a black-box method, \probtest only requires access to execute the system.
As a testing method, \probtest can impose notably lower requirements in the number of executions compared to SMC.
For example, SMC can be used to estimate an assertion violation probability up to some error \(\delta\) and with probability \(\epsilon\), formally \(P(p-\delta \leq p \leq p+\delta) \geq 1 - \epsilon\).
A common statistical bound to tackle this problem (without a high requirement in the number of executions) is the Hoeffding bound, which states that \( \delta = \sqrt{\ln(2/\epsilon)/2k} \) where $k$ is the number of executions~\cite{meggendorfer2025oddsimprovingfoundationsstatistical}.
Let \(\delta = 0.01\) and \(\epsilon = 0.05\), then this bound requires \(> 70000\) executions.
This would be the required number of executions for the case studies in~\cref{chapter:evaluation}, while \probtest\ required less than \(3000\) executions, i.e., an order of magnitude less.
This is expected, as SMC estimates the assertion violation probability whereas \probtest checks whether a given one holds.
Nevertheless,  it is noteworthy that, in this setup, \probtest can ensure statistical guarantees with notably less executions.

Hypothesis testing has been used to statistically test properties of randomized algorithms~\cite{DBLP:conf/icse/ArcuriB11}.
A set of guidelines to test whether two different randomized algorithms exhibit the same statistical behavior (e.g., expectation or variance) was proposed in~\cite{DBLP:conf/icse/ArcuriB11}.
A black-box testing method for probabilistic programs based on hypothesis testing method has been proposed in~\cite{testing_pp_statistical_hypothesis}.
The method allows testers to check whether the distribution of a program's output has a given expectation and variance.
This allows to test more general properties than in \probtest, as there are no constraints on the program output distribution.
However, the method assumes that testers can specify a program output distribution, which may require more advanced proficiency in statistics compared to \probtest.
The method can be used to check whether an assertion holds with certain probability by using a Bernoulli distribution with mean \(1-p\) where \(p\) is the assertion violation probability.
But here again the required number of executions may be larger than for \probtest.
This method requires executing the program \(k = (2\lceil -24\ln(\epsilon) \rceil + 1)n\) (with \(n>31\)) to test whether that an assertion violation probability \(p\) holds with probability \(1-\epsilon\)~\cite{testing_pp_statistical_hypothesis}.
Consider \(\epsilon=0.05\) as in the case studies in~\cref{chapter:evaluation}, then the program must be executed (at least) 4433 times.
This is \(\approx1000\) times more than the largest \(k\) for \probtest in our case studies. 
It is also noteworthy that for many assertion violation probabilities \probtest required less than \(1000\) executions, which is 4 times less than this hypothesis testing method.
Using the same statistical test as in~\cite{testing_pp_statistical_hypothesis}, a method for metamorphic testing has been proposed~\cite{DBLP:conf/qsic/GuderleiM07}.
This method does not require specifying the exact program output distribution, but requires specifying statistical metamorphic relations between inputs and outputs.

Dutta et al. propose a method to test probabilistic programming frameworks~\cite{DBLP:conf/sigsoft/DuttaLHM18}---these are frameworks to define Bayesian probabilistic models (e.g., \cite{pymc,pyro,stan}).
The method works by checking equivalence between inferred posterior distributions and analytical solutions (computed by tools like PSI~\cite{PSI}) or other approximations (computed using probabilistic inference methods such as MCMC).
Although, this method could be applied in our setting, it is not designed for this purpose.
Finally, this is a white-box method, as opposed to our method that only requires black-box access to the program.
In \cite{varshosaz2023}, a test harness is proposed for reinforcement learning, including also statistical tests tuned by trial and error, underscoring the need for systematic testing for such applications.
\looseness -1

The coupon collector's problem has been applied to analyse the effectiveness and predictability of random testing \cite{ccp_randomtesting} and scenario-based testing of autonomous driving systems~\cite{DBLP:conf/itsc/HauerSHP19}. 
The source of randomness in this line of work is on generating program inputs, but not the probabilistic behavior of the program under test.
\probtest could be combined with random testing to support property-based testing of probabilistic programs.


\section{Conclusion}

We have presented, \probtest, a novel method for unit testing probabilistic programs in a black-box setting. 
Given an assertion specification (i.e., the probability for the probabilistic program to violate/satisfy the assertion), we have proven that \probtest achieves assertion coverage with high probability.
We have studied the scalability of our method as the correctness requirements increase, and its sensitivity to inaccurate assertion violation specifications. 
We have evaluated the effectiveness of \probtest to find bugs in two case studies: 
\begin{inparaenum}[i)]
\item a randomized data structure, the skip list, which is often used in industrial applications; and
\item two RL problems from the Gymnasium Python library, which is a popular library for implementing RL agents.
\end{inparaenum}
Our results demonstrate that \probtest is an effective method to test probabilistic programs, and also it is applicable to realistic software systems.


\bibliographystyle{plain}
\bibliography{refs}

\newpage
\appendix
\section{Properties of a skip list} \label{appendix:skip_list_properties}

Here we list each of the 26 tests that we used to test the skip list implementation.
Each test is described as a property using a syntax similar to that of Hoare triples; second column in the table below.
That is, \(\{P\}\texttt{test\_body}\{Q\}\) where \(P,Q\) are the pre- and post-conditions of the test and \texttt{test\_body} is the body of the test.
Note that the body of the test does not contain the assertion.
The assertion is model by the post-condition \(Q\).
Furthermore, we specify the name of the corresponding test in the accompanying artifact; third column in the table below.
The first column of the table specifies a bug ID, and a header indicates the bug group: validity, post-conditions, metamorphic or equivalence.

Let $L$ be the set of all valid skip lists and $N$ be the set of all nodes.
Consider the two equivalence relations on skip lists:
\begin{itemize}
    \item[] \(=\): Equality between skip lists, i.e., contains the same nodes and all pointers the same.
    \item[] $\equiv$: Contains the same nodes, but may be different in structure.
\end{itemize}
There are some auxiliary methods below (such as {\tt \_levels\_of\_nodes}) that are not included in our implementation. We use them here to simplify notation and improve readability.

\small

\begin{longtable}{| c | >{\raggedright \ttfamily}m{20em} | >{\ttfamily}m{15em} |}
  \hline
  
  \multicolumn{3}{|c|}{Validity} \\ \hline
  Id & \text{Property} & Test \\ \hline
  1  & $\{l\in L \land n \in N\}$ l.insert(n) $\{l \in L\}$ & test\_valid\_insertion \\ 
  \hline
  2 & $\{l\in L \land n \in N\}$ l.delete(n) $\{l \in L\}$ & test\_valid\_delete \\ 
  \hline
  3 &  $\{l\in L \land n_1,...,n_m \in N\}$ l.delete(n\_1,...,n\_m) $\{l \in L\}$ & test\_valid\_delete\_all 
  \\ 
  \hline
  4 & $\{l\in L \land \in N\}$ l.search(n) $\{l \in L\}$ & test\_valid\_search

  \\ \hline \hline

    \multicolumn{3}{|c|}{Postconditions} \\ \hline
    Id & Property & Test \\ \hline

  5  & $\{l\in L \land n \in N \land n \in l\}$ l.contains(n) = c; l.search(n).key = v \(\{c \implies n.\mathit{key} = v\}\) &  test\_post\_insertion\_search \\ 
  \hline
  6 & $\{l\in L \land n \in N\}$ l.delete(n); l.search(n) = v \(\{v = \texttt{None}\}\) & test\_post\_deletion\_search \\ 
  \hline
  7 & $\{l\in L \land n_1,...,n_m \in N\}$ v\_1 = l.\_levels\_of\_nodes(n\_1,...,n\_m-1); l.insert(n\_m); v\_2 = l.\_levels\_of\_nodes(n\_1,...,n\_m-1) \(\{v_1 = v_2\}\) & test\_post\_level\_of\_nodes\_ unchanged\_after\_insertion \\ 
    \hline
  8 & $\{l\in L \land n_1,...,n_m \in N \land i \in \{1,...,m\}\}$ v\_1 = l.\_levels\_of\_nodes(n\_1,...,n\_m), l.search(n\_i);  v\_2 = l.\_levels\_of\_nodes(n\_1,...,n\_m) \(\{v_1 = v_2\}\) & test\_post\_level\_of\_nodes\_ unchanged\_after\_search \\ 
    \hline
  9 & $\{l\in L \land n_1,...,n_m \in N \land i \in \{1,...,m\}\}$ v\_1 = l.\_levels\_of\_nodes(n\_1,...,n\_i-1,n\_i+1, ...,n\_m); l.delete(n\_i); v\_2 =  l.\_levels\_of\_nodes(n\_1,...,n\_i-1,n\_i+1
  ,...,n\_m) \(\{v_1 = v_2\}\) & test\_post\_level\_of\_other\_ nodes\_unchanged\_after\_ deletion \\ 
      \hline
  10 & $\{l\in L \land n \in N \land n \in l\}$ v\_1 = l.level\_of\_node(n); l.delete(n); v\_2 = l.level\_of\_node(n) \(\{v_1 \not= v_2\}\) & test\_post\_level\_of\_deleted\_ nodes\_changed\_after\_deletion 

  \\ \hline \hline

    \multicolumn{3}{|c|}{Metamorphic properties} \\ \hline
    Id & Property & Test \\ \hline

    11 & $\{l_1,l_2\in L \land n_1,n_2 \in N \land l_1 \equiv l_2\}$  l\_1.insert(n\_1); l\_1.insert(n\_2); l\_2.insert(n\_2); l\_2.insert(n\_1) \(\{l_1 \equiv l_2\}\) & test\_meta\_insertion\_order  \\ \hline
    12 & $\{l_1,l_2\in L \land n \in N \land l_1 = l_2 \land n \in l_1\}$  l\_1.insert(n) \(\{l_1 = l_2\}\) & test\_meta\_insertion\_order\_ same\_key \\ \hline
    13 & $\{l_1,l_2\in L \land n_1,n_2 \in N \land l_1 \equiv l_2\}$ l\_1.insert(n\_1); l\_1.delete(n\_1); l\_1.insert(n\_2); l\_2.insert(n\_1); l\_2.insert(n\_2); l\_2.delete(n\_2) \(\{l_1 \equiv l_2\}\) & test\_meta\_delete\_insert\_order \\ \hline
    14 & $\{l_1,l_2\in L \land n \in N \land l_1 \equiv l_2 \land n \in l_1\}$ l\_1.insert(n); l\_2.delete(n); l\_2.insert(n) \(\{l_1 \equiv l_2\}\) & test\_meta\_insert\_delete\_ same\_node\_order \\ \hline
    15 & $\{l_1,l_2\in L \land n_1,n_2 \in N \land l_1 = l_2\}$ l\_1.delete(n\_1); l\_1.delete(n\_2); l\_2.delete(n\_2); l\_2.delete(n\_1) \(\{l_1 = l_2\}\) & test\_meta\_delete\_delete\_order \\ \hline
    16 & $\{l_1,l_2\in L \land n \in N \land l_1 = l_2\}$  l\_1.delete(n); l\_1.delete(n); l\_2.delete(n) \{\(l_1 = l_2\)\} & test\_meta\_delete\_same\_ node\_twice \\ \hline
    17 & $\{ l_1,l_2\in L \land n \in N \land l_1 = l_2 \land n \in l_1\}$  l\_2.delete(n); v\_1 = l\_1.search(n); v\_2 = l\_2.search(n) \(\{v_1 \not= v_2\}\)& test\_meta\_insertion\_search\_ delete \\ \hline
    18 & $\{l_1,l_2\in L \land n \in N \land l_1 = l_2 \land n \not \in l_1 \}$  l\_1.insert(n); l\_1.delete(n); v\_1 = l\_1.search(n); v\_2 = l\_2.search(n) \(\{v_1 = v_2\}\) & test\_meta\_search\_insertion\_ delete 

      \\ \hline \hline

    \multicolumn{3}{|c|}{Properties on equivalence} \\ \hline
    Id & Property & Test \\ \hline
    19 & $\{l_1,l_2\in L \land \forall n \in N \land l_1 = l_2\}$  l\_1.insert(n) \(\{l_1 \not= l_2\}\) & test\_eq\_inserts\_break\_eq \\ \hline
    20 & $\{l_1,l_2\in L \land \forall n \in N \land l_1 = l_2\}$  l\_1.delete(n) \(\{l_1 \not= l_2\}\) & test\_eq\_delete\_break\_eq\\ \hline
    21 & $\{l_1,l_2\in L \land \forall n \in N \land l_1 = l_2\}$  l\_1.delete(n); l\_2.delete(n) \(\{l_1 = l_2\}\) & test\_eq\_delete\\ \hline
    22 & $\{l_1,l_2\in L \land \forall n \in N \land l_1 = l_2\}$  l\_1.search(n) \(\{l_1 = l_2\}\) & test\_eq\_search \\ \hline
    23 & $\{l_1,l_2\in L \land n \in N \land l_1 \equiv l_2\}$  l\_1.insert(n); l\_2.insert(n) \(\{l_1 \equiv l_2\}\) & test\_equiv\_insert \\ \hline
    24 & $\{l_1,l_2\in L \land n \in N \land l_1 \equiv l_2\}$  l\_1.insert(n) \(\{l_1 \not \equiv l_2\}\)  & test\_equiv\_inserts\_break\_ equiv \\ \hline
    25 & $\{l_1,l_2\in L \land n \in N \land l_1 \equiv l_2\}$  l\_1.delete(n) \(\{l_1 \not \equiv l_2\}\) & test\_equiv\_delete\_break\_equiv\\ \hline
    26 & $\{l_1,l_2\in L \land n \in N \land l_1 \equiv l_2\}$  l\_1.delete(n); l\_2.delete(n) \(\{l_1 \equiv l_2\}\) & test\_equiv\_delete \\ \hline

\caption{Properties under consideration for testing the skip list, and the corresponding test name in the test suite for each property.}
\label{table:skiplist_properties}
\end{longtable}
\normalsize

\section{Bugs injected in the skip list} \label{appendix:skiplist_bugs}

Here we describe each of the 8 bugs that we manually injected in our skip list implementation.
Each row in the table below shows contains: the bug ID, a description of the bug behavior, an example showcasing the effect of the bug, and a note indicating whether the bug is detectable in one or more of the resulting skip lists after executing \(R \in {1,2,3,4}\) insertions.
Table headers indicate the minimum number of insertions \(R\) required to trigger the bug.
In the examples, we use \(n_1 \to n_2\) or \(n \to \texttt{None}\) to denote that the \texttt{next} pointer of \(n_1\) points to \(n_2\) or \texttt{None}.
The link list associated to a level \(L_i\) in the skip list is denoted as \(n_1 \to \ldots \to n_m \to \texttt{None}\).
We write the different levels of a skip lists in different lines.
For instance, \\
$2 \to$ {\tt None} \\
$1 \to 2 \to$ {\tt None} \\
denotes a skip list with 2 layers where \(L_1 = 1 \to 2 \to \texttt{None}\) and \(L_2 = 2 \to \texttt{None}\).

\small
\begin{longtable}{| c | >{\raggedright\arraybackslash}m{8em} | p{15em} | >{\raggedright\arraybackslash}m{7.5em} |}
  \hline

\multicolumn{4}{|c|}{Bugs for $R\geq 1$} \\ \hline
Bug id & Description & Example & Notes \\

\hline

1 & When deleting a node at the top layer $L_{M}$, nothing happens. & 
\makecell[l]{
Given the skip list: \\
$12 \to$ {\tt None} \\
$12 \to$ {\tt None} \\
$12 \to$ {\tt None}\\ \\
Delete node 12: \\ \\
$12 \to$ {\tt None} \\
$12 \to$ {\tt None} \\
$12 \to$ {\tt None}
}
& For $R=1$ occurs in a single outcome. \\

\hline

2 & When raising a node, its corresponding node in the bottom layer $L_1$ gets a lower pointer to a node. & 
\makecell[l]{
Given the skip list: \\
$12 \to$ {\tt None} \\
$12 \to$ {\tt None} \\ \\
\texttt{str(l.nodes\_at\_level(0)[0])} \\
returns \\ \texttt{12 (next: None, lower: 12)}
}

& Affects several outcomes \\

\hline
\hline

\multicolumn{4}{|c|}{Bugs for $R\geq 2$} \\ \hline
Bug id & Description & Example & Notes \\
\hline

3 & When two or more nodes are raised to $L_M$, this layer is deleted/emptied. & 
\makecell[l]{
Given the skip list: \\
$15 \to$ {\tt None} \\
$15 \to$ {\tt None} \\
$15 \to$ {\tt None} \\ \\
Insert 16 might result in: \\ \\
$15 \to 16 \to$ {\tt None} \\ 
$15 \to 16 \to$ {\tt None}
}
& For $R=2$ occurs in a single outcome. \\

\hline

4 & When two or more nodes have been raised to layers above $L_1$, searching for a raised node decrements the found node's key by 1.
& 
\makecell[l]{
Given the skip list: \\
$15 \to 16 \to$ {\tt None} \\
$15 \to 16 \to$ {\tt None} \\ \\
search(16) returns the node \\ \texttt{15 (next: None, lower: 16)} \\ and changes the list to: \\ \\
$15 \to 15 \to$ {\tt None} \\
$15 \to 16 \to$ {\tt None}
}
& Affects several outcomes \\

\hline
\hline

\multicolumn{4}{|c|}{Bugs for $R\geq 3$} \\ \hline
Bug id & Description & Example & Notes \\
\hline

5 & When three nodes or more have been raised to the maximum level $L_m$, the level of these nodes will return 0. & 
\makecell[l]{
Given the skip list: \\
$13 \to 15 \to 16 \to $ {\tt None} \\
$13 \to 15 \to 16 \to $ {\tt None} \\
$13 \to 15 \to 16 \to $ {\tt None} \\ \\
\texttt{level\_of\_node(15)} returns 0
}
& For $R=3$ occurs in a single outcome. \\

\hline

6 & If three nodes or more have the same level, they all get raised.  & 
\makecell[l]{
Given the skip list:\\
$13 \to 16 \to $ {\tt None} \\
$13 \to 16 \to $ {\tt None} \\ \\
insert(15) might result in \\ \\
$13 \to 15 \to 16 \to $ {\tt None} \\
$13 \to 15 \to 16 \to $ {\tt None} \\
$13 \to 15 \to 16 \to $ {\tt None} \\ \\
}
& Affects several outcomes \\

\hline
\hline

\multicolumn{4}{|c|}{Bugs for $R\geq 4$} \\ \hline
Bug id & Description & Example & Notes \\
\hline

7 & When four or more nodes have been raised to $L_M$, searching for any of these nodes returns its next pointer. & 
\makecell[l]{
Given the skip list: \\
$12 \to 13 \to 15 \to 16 \to$ {\tt None} \\
$12 \to 13 \to 15 \to 16 \to$ {\tt None} \\
$12 \to 13 \to 15 \to 16 \to$ {\tt None} \\ \\
search(13) returns {\tt None}
}
& For $R=4$ occurs in a single outcome. \\

\hline

8 & Searching for a node that has level $>1$ and is at the head in a skip list containing 4 or more nodes returns {\tt None}. & 
\makecell[l]{
Given the skip list: \\
$13 \to$ {\tt None} \\
$12 \to 13 \to 15 \to 16 \to$ {\tt None} \\ \\
search(13) returns {\tt None}
}
& Affects several outcomes \\

\hline

\caption{Bugs injected into the skip list implementation.}
\label{table:skiplist_injected_bugs}
\end{longtable}

\normalsize

\section{Additional results for RL case studies}
\label{appendix:more_rl_results}

\Cref{tbl:table_complete_rl_experiments} shows the complete list of experiments conducted for RL case studies.
The probability of property failure (column 5) is estimated by evaluating the trained policy over 100k executions.
The probability of \probtest finding a bug (column 6) is estimated by running \probtest 100 times for the required number of executions.
The most important remark of these results is that \probtest finds failures with probability \(0.99\) for all cases; despite \(\epsilon=0.05\) only requiring \(\geq 0.95\).

\begin{table}[t!]
\centering
\scalebox{1.0}{
  \begin{tabular}{
  c@{\hskip 4mm}
  c@{\hskip 4mm}
  c@{\hskip 4mm}
  c@{\hskip 4mm}
  c@{\hskip 4mm}
  c@{\hskip 4mm}
}
\toprule
experiment & p & property & \makecell{Num.\\episodes} & \makecell{Prob. property\\does not hold} & \makecell{Prob. Probtest\\finds bug} \\
\midrule
4x4 & 0.01 & Q1 & 10 & 0.530 & 1.000 \\
4x4 & 0.01 & Q1 & 20 & 0.119 & 1.000 \\
4x4 & 0.01 & Q1 & 30 & 0.016 & 1.000 \\
4x4 & 0.01 & Q1 & 40 & 0.000 & 0.000 \\
4x4 & 0.01 & Q1 & 50 & 0.000 & 0.000 \\
4x4 & 0.01 & Q1 & 60 & 0.000 & 0.000 \\
4x4 & 0.01 & Q1 & 70 & 0.000 & 0.000 \\
4x4 & 0.01 & Q1 & 80 & 0.000 & 0.000 \\
4x4 & 0.01 & Q1 & 90 & 0.000 & 0.000 \\
4x4 & 0.01 & Q1 & 100 & 0.000 & 0.000 \\
4x4 & 0.001 & Q2 & 10 & 0.011 & 1.000 \\
4x4 & 0.001 & Q2 & 20 & 0.095 & 1.000 \\
4x4 & 0.001 & Q2 & 30 & 0.715 & 1.000 \\
4x4 & 0.001 & Q2 & 40 & 0.226 & 1.000 \\
4x4 & 0.001 & Q2 & 50 & 0.002 & 1.000 \\
4x4 & 0.001 & Q2 & 60 & 0.002 & 0.990 \\
4x4 & 0.001 & Q2 & 70 & 0.002 & 0.990 \\
4x4 & 0.001 & Q2 & 80 & 0.001 & 0.990 \\
4x4 & 0.001 & Q2 & 90 & 0.002 & 1.000 \\
4x4 & 0.001 & Q2 & 100 & 0.002 & 0.990 \\
7x7 & 0.01 & Q1 & 10000 & 0.055 & 1.000 \\
7x7 & 0.01 & Q1 & 15000 & 0.012 & 0.990 \\
7x7 & 0.01 & Q1 & 20000 & 0.000 & 0.000 \\
7x7 & 0.01 & Q1 & 25000 & 0.000 & 0.000 \\
7x7 & 0.001 & Q2 & 10000 & 0.082 & 1.000 \\
7x7 & 0.001 & Q2 & 15000 & 0.051 & 1.000 \\
7x7 & 0.001 & Q2 & 20000 & 0.082 & 1.000 \\
7x7 & 0.001 & Q2 & 25000 & 0.060 & 1.000 \\
cliffwalk & 0.001 & Q & 15000 & 0.007 & 1.000 \\
cliffwalk & 0.001 & Q & 20000 & 0.007 & 1.000 \\
cliffwalk & 0.001 & Q & 25000 & 0.007 & 1.000 \\
\bottomrule
\end{tabular}

}
\caption{List of experiment results for RL case studies.}
\label{tbl:table_complete_rl_experiments}
\end{table}


\end{document}